\documentclass[twocolumn,floats,floatfix,prb,aps, superscriptaddress,usenames,dvipsnames]{revtex4-2}
 
\usepackage{blindtext}

\usepackage{amsmath}
\usepackage{amssymb}
\usepackage{floatrow}
\usepackage{dcolumn}
\usepackage{graphicx}
\usepackage[caption=false, label font=bf,labelformat=simple]{subfig}

\usepackage[usenames,dvipsnames]{color}
\usepackage[table]{xcolor}
\usepackage{multirow}
\usepackage{bm} 
\usepackage{siunitx} 
\usepackage{physics} 
\usepackage[normalem]{ulem}
\usepackage{soul}
\usepackage[super]{nth}

\usepackage{placeins}

\DeclareSIUnit{\Ry}{Ry}
\DeclareSIUnit{Å}{Åstrom}
\DeclareSIUnit{\G}{\text{\ensuremath{G_0}}}
\DeclareSIUnit{\Debye}{D}
\DeclareSIUnit{\muB}{\text{\ensuremath{\mu_B}}}

\newcommand{\mr}[1]{\multirow{2}{*}{#1}}  
\newcommand{\Z}{\ensuremath{\mathbb{Z}_2}}
\newcommand{\ttimes}{\ensuremath{\setlength{\medmuskip}{0mu}\times}}
\newcommand{\RotatedUC}{(\ensuremath{\sqrt{3}\ttimes\sqrt{3})}R30$^{\circ}$}

\usepackage{standalone}

\begin{document}

\title{Tuning the Topological Band Gap of Bismuthene with Silicon-based Substrate}

\author{Nils Wittemeier}
\affiliation{Catalan Institute of Nanoscience and Nanotechnology - ICN2 (CSIC and BIST), Campus UAB, Bellaterra, 08193 Barcelona, Spain}


\author{Pablo Ordej\'on}
\affiliation{Catalan Institute of Nanoscience and Nanotechnology - ICN2 (CSIC and BIST), Campus UAB, Bellaterra, 08193 Barcelona, Spain}

\author{Zeila Zanolli}
\affiliation{Chemistry Department and, Debye Institute for Nanomaterials Science, Condensed Matter and Interfaces, Utrecht University and European Theoretical Spectroscopy Facility, PO Box 80.000, 3508 TA Utrecht, The Netherlands}


\date{\today}

\begin{abstract}
Some meta-stable polymorphs of bismuth monolayer (bismuthene) can host topologically nontrivial phases. However, it remains unclear if these polymorphs can become stable through interaction with a substrate, whether their topological properties are preserved, and how to design an optimal substrate to make the topological phase more robust. 
Using first-principles techniques we demonstrate that bismuthene polymorphs can become stable over silicon carbide (SiC), silicon (Si), silicon dioxide (SiO2) and that the proximity interaction in the heterostructures has a significant effect on the electronic structure of the monolayer, even when bonding is weak. We show that the van der Waals interactions and the breaking of the sublattice symmetry are the main factors driving changes in the electronic structure. Our work demonstrates that substrate interaction can strengthen the topological properties of bismuthene polymorphs and make them accessible for experimental investigation and technological applications.
\end{abstract}

\maketitle

\section{Introduction}


Topological insulators (TI) are among the most promising candidates for next-generation nanoscale electronic devices due to their ability to conduct currents through edges states without dissipation
\,\cite{hasan_colloquium_2010,qi_topological_2011,moore_birth_2010}. 
However, the ability to generate robust TI and to control their band gap remains a critical issue for applications.
Elements with strong intrinsic spin-orbit coupling (SOC), like Bismuth, are likely to be good TI candidates.
Atomically thin Bi(111) layers are topological\,\cite{koroteev_first-principles_2008,liu_stable_2011,ma_two-dimensional_2015,zhou_formation_2015,singh_low-energy_2019}
but present an electronic band gap (0.08\,eV)\,\cite{singh_low-energy_2019} too small for room-temperature operation.
A possible solution is to find a substrate that 
enhances the gap while preserving the topological properties.
%
Substrates also can induce a TI phase in topologically trivial monolayers:
flat-hexagonal (f-hex) bismuth monolayer is trivial in the free-standing form but becomes topological when grown on SiC(0001)\,\cite{reis_bismuthene_2017}, or partially passivated Si(111)\,\cite{zhou_formation_2015}.

Singh et al.\,\cite{singh_low-energy_2019} recently predicted that there are
other Bi monolayer phases besides the ground state puckered monoclinic (Bi(110)): 
buckled hexagonal (b-hex, Bi(111)), $\alpha$, $\beta$, $\gamma$, f-hex, in order of formation energy. 
In free-standing form only the b-hex and $\gamma$ phases are topologically non-trivial ~\cite{singh_low-energy_2019}.
The interaction with a suitable substrate could stabilize these meta-stable phases.
However, only the b-hex and f-hex present the in-plane hexagonal symmetry
as graphene and, therefore, can be Haldane-type TIs.
%
%

In this paper, we explore the effect of silicon carbide (SiC), silicon (Si), and silicon dioxide (SiO$_2$) substrates on the topological features of the two hexagonal phases 
of bismuthene.
Thin layers of Bi have been synthesized on SiC\,\cite{reis_bismuthene_2017} and Si\,\cite{kawakami_one-dimensional_2015}. 
The SiO$_2$ substrate was instrumental for the experimental observation of mechanically exfoliated graphene\,\cite{novoselov_electric_2004} and has been widely used as a dielectric medium in integrated circuits.
We demonstrate that only the buckled hexagonal Bi retains its topological phase on all surfaces we considered, while f-hex bismuthene undergoes a trivial to topological phase transition only when placed on top of SiC.
We further demonstrate that the interaction between b-hex and SiO$_2$ is highly dependent on the surface structure: a hydroxylated SiO$_2$ surface can enhance the topological band gap, and a cleaved silicon terminated surface can close the band gap significantly.

\section{Methods}
We carry out simulations using density functional theory (DFT) as implemented in the SIESTA code\,\cite{soler_siesta_2002,garcia_siesta_2020}, including spin-orbit coupling in the fully relativistic pseudopotential formalism\,\cite{cuadrado_fully_2012, cuadrado_validity_2021}. We employ fully relativistic, optimized norm-conserving Vanderbilt pseudopotentials (ONCVPSP\,\cite{hamann_optimized_2013}) in PSML format\,\cite{garcia_psml_2018} from PseudoDojo database\,\cite{van_setten_pseudodojo_2018} generated with PBE\,\cite{perdew_generalized_1996} exchange-correlation functional. The Kohn-Sham equations are solved using standard double-zeta polarized basis sets, a real space grid with a cut-off of 600\,Ry, an electronic temperature of 5\,meV and a $15\ttimes 15\ttimes 1$ Monkhorst-Pack grid for a single bismuthene unit cell. For larger cells the k point sample is scaled accordingly. All structures are relaxed with a force threshold of 0.01\,eV\/Å and a maximum stress tolerance of 0.006\,eV\/Å$^3$.

For the isolated substrate slabs, we employ the PBE functional~\cite{perdew_generalized_1996}, and for the heterostructures the van der Waals density functional of M. Dion et al.\,\cite{dion_van_2004} (vdW-DRSSL). We employ PBE and vdW-DRSSL functional both for the free-standing bismuthene phases and compare the difference in the resulting electronic structure.
We ensure decoupling of the upper and lower slab surfaces by imposing a 1\,meV convergence threshold on the change in surface energy for any additional layer. We use a dipole correction to prevent exaggeration of the slab dipole due to interaction between periodic images.
The \Z\,topological invariant is determined using hybrid Wannier charge centres with Z2Pack\,\cite{gresch_z2pack_2017}.

We fix the lattice vectors of the substrate, and strain the bismuthene phase. This approach recreates experimental conditions in which the monolayer will adjust to the more rigid substrate. We apply a counterpoise correction for the basis set superposition error\,\cite{boys_calculation_1970} to evaluate the binding energies ($E_B$) between monolayer and substrate. 

\begin{table*}
      \definecolor{lightgray}{gray}{0.94}
      \centering
      \setlength{\tabcolsep}{8pt}
      \renewcommand{\arraystretch}{1.3}
      \begin{tabular}{l l l c c c c c} 
            \hline 
        \mr{Bi phase}& \mr{Substrate}& & \mr{\Z} & Lattice        & $E_B$ per Bi [eV] & Interlayer     & $E_g[eV]$\\
                     &               & &         & constant [\AA] &                   & distance [\AA] & \\
            \hline 
            b-hex    &\textit{freestanding} &                & 1 & 4.28 & --   & --   & 0.38\\
                     & SiC(0001)     & Si term. \& H passiv. & 1 & 3.12 & 0.08 & 2.73 & 0.18 \\
                     & Si(111)       & H passivated          & 1 & 3.88 & 0.07 & 2.66 & 0.12\\
                     & SiO$_2$(0001) & hydroxylated          & 1 & 5.03 & 0.06 & 2.90 & 0.40\\
                     & SiO$_2$(0001) & reconstructed         & 1 & 5.03 & 0.07 & 2.96 & 0.38\\
                     & SiO$_2$(0001) & Si terminated         & 1 & 5.03 & 0.11 & 2.34 & 0.08\\
            \hline 
            f-hex    &\textit{freestanding} &                & 0 & 5.35 & --   & --   & 0.51\\
                     & SiC(0001)     & Si terminated         & 1 & 3.12 & 1.49 & 2.76 & 0.62\\
                     & SiO$_2$(0001) & hydroxylated          & 0 & 5.03 & 0.11 & 3.22 & 0.19\\
                     & SiO$_2$(0001) & reconstructed         & 0 & 5.03 & 0.09 & 2.76 & 0.80\\
                     & SiO$_2$(0001) & Si terminated         & 0 & 5.03 & 0.19 & --   & --\\
            \hline 
      \end{tabular}
      \caption{Summary of basic properties of   buckled-hexagonal and flat-hexagonal Bi monolayer phases in freestanding format and supported by different substrates.}
      \label{tab:summary}
\end{table*}

\section{Results and Discussion}

\subsection{Free-standing bismuthene}
\begin{figure}
    \centering%
    \begin{minipage}[t]{.44\textwidth}
        \subfloat[]{
            \includegraphics[trim={25cm 10cm 24cm 4cm},clip,width=0.94\textwidth]{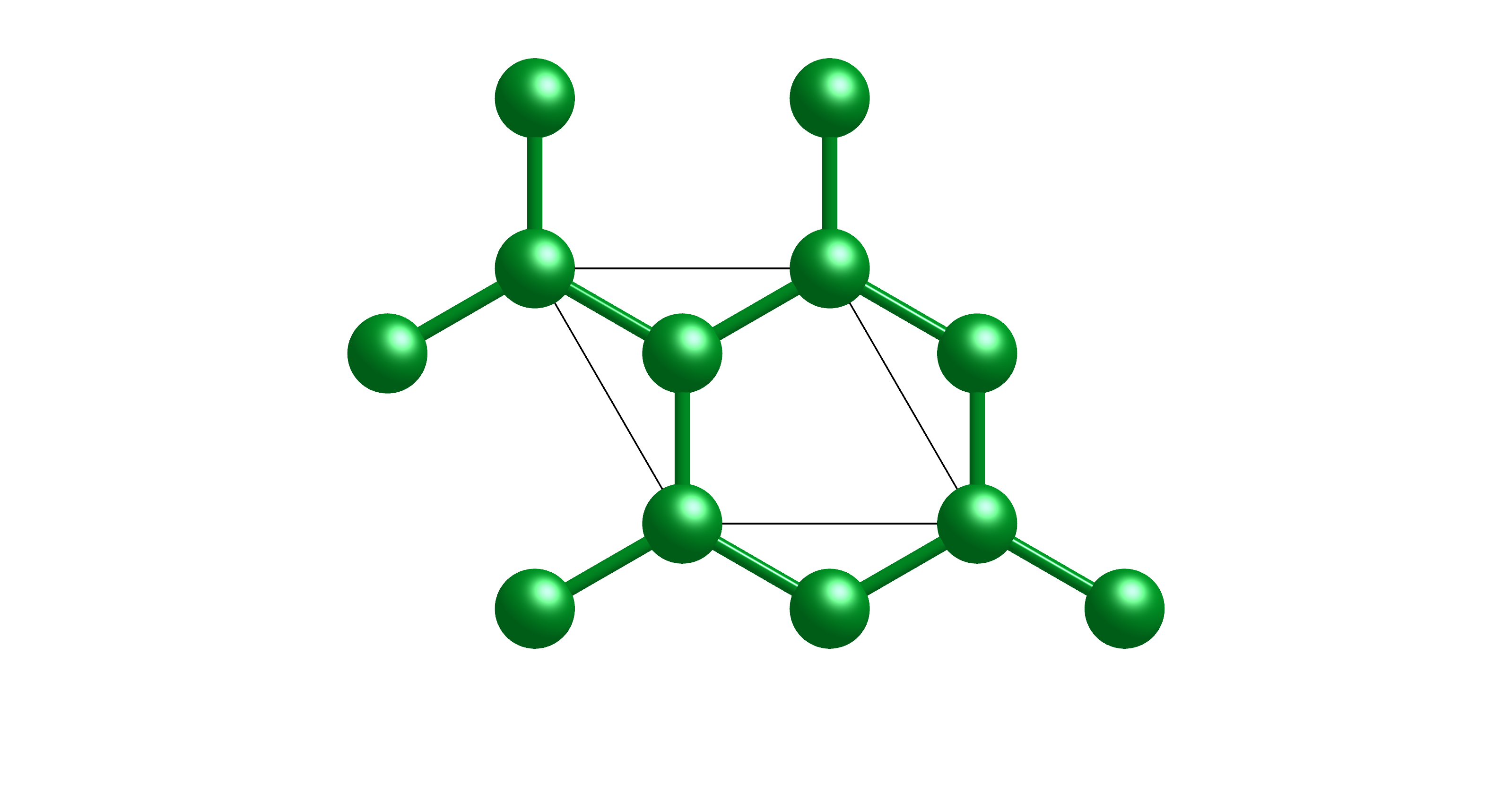}
            \label{fig:f-hex_struct_top}
        }
        
        \subfloat[]{
            \includegraphics[trim={20cm 10cm 20cm 7cm},clip,width=0.94\textwidth]{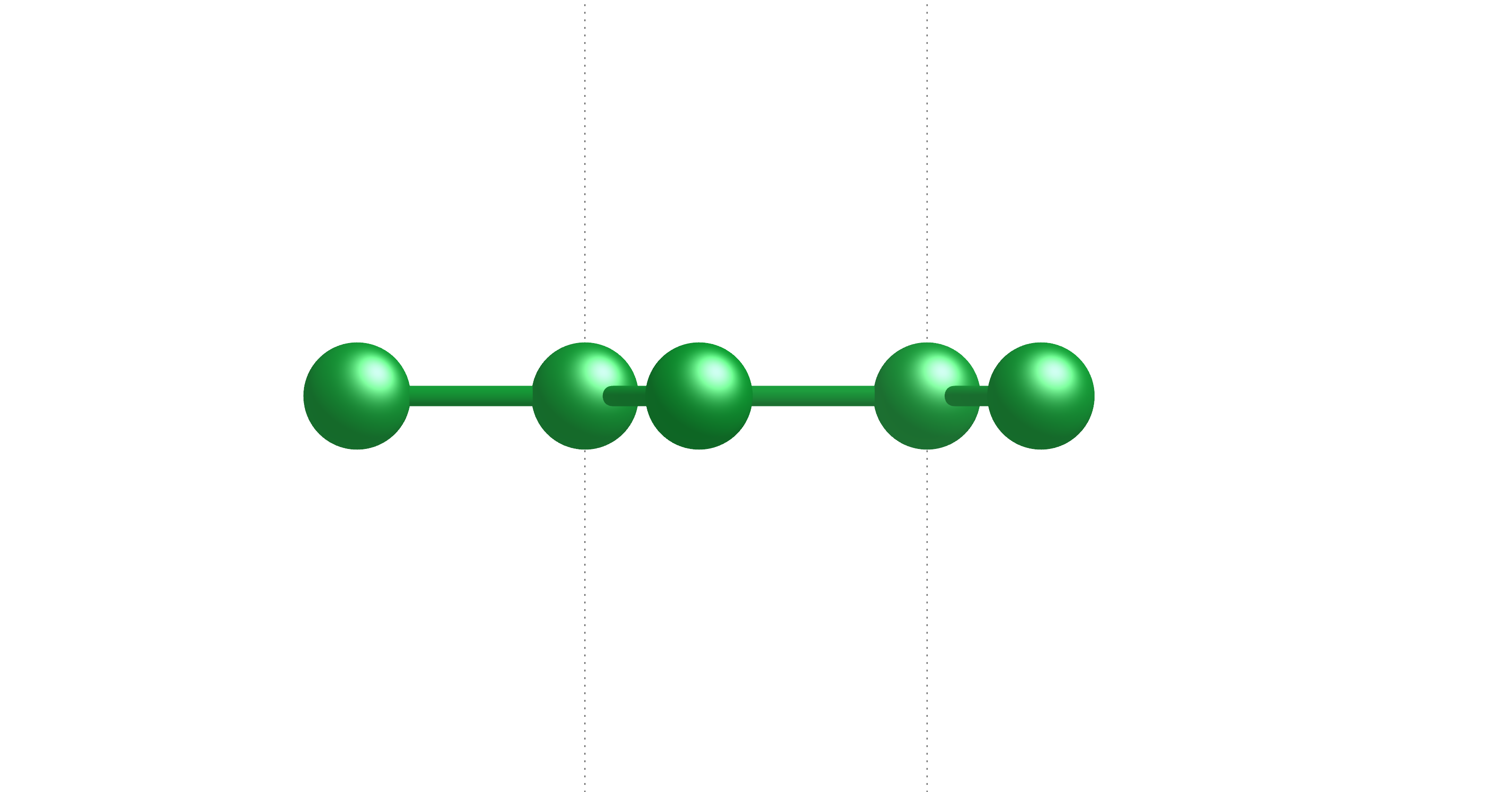}
            \label{fig:f-hex_struct_side}
        }
    \end{minipage}
    \begin{minipage}[t]{.50\textwidth}
        \subfloat[]{
            \includegraphics[width=0.94\textwidth]{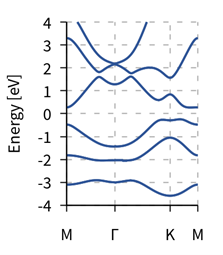}
            \label{fig:f-hex_bands}
        }
    \end{minipage}
    
    \begin{minipage}[t]{.40\textwidth}
        \subfloat[]{%
            \includegraphics[trim={29cm 10cm 29cm 7cm},clip,width=0.94\linewidth]{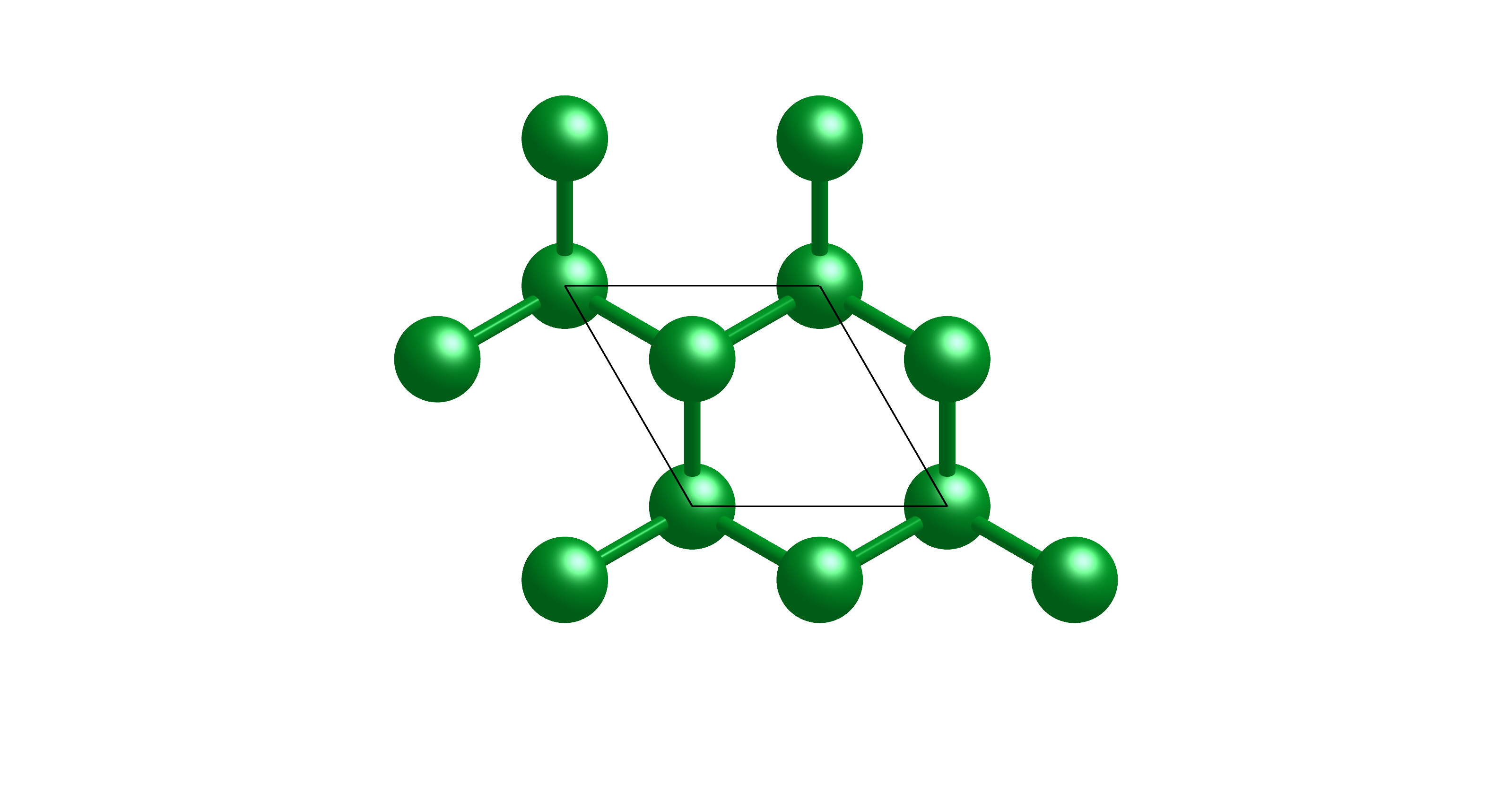}%
            \label{fig:b-hex_struct_top}%
        }
        
        \subfloat[]{%
            \includegraphics[trim={28cm 15cm 24cm 20cm},clip,width=0.94\textwidth]{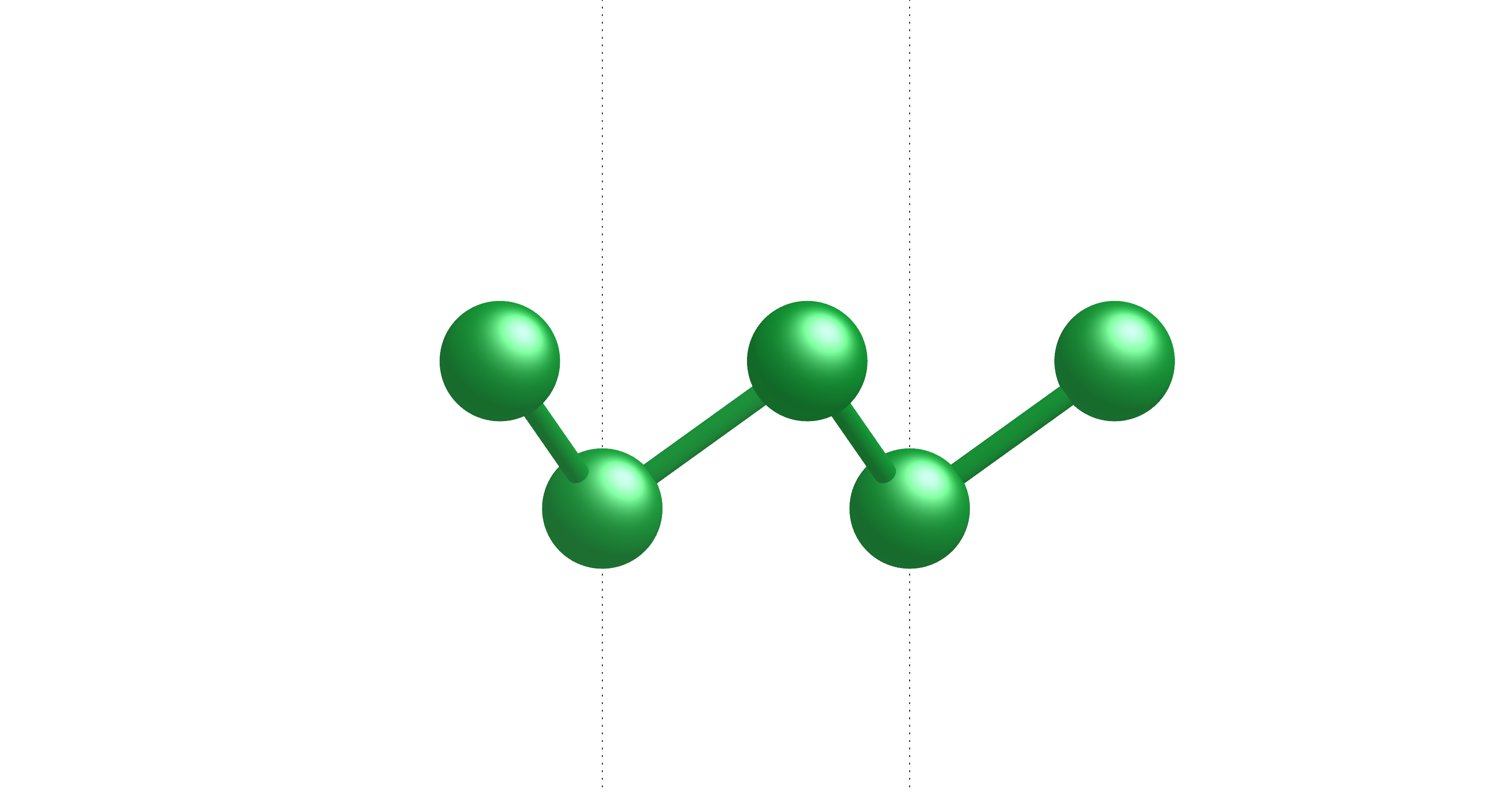}%
            \label{fig:b-hex_struct_side}%
        }
    \end{minipage}
    \begin{minipage}[t]{.54\textwidth}
        \subfloat[]{%
            \includegraphics[width=0.94\textwidth]{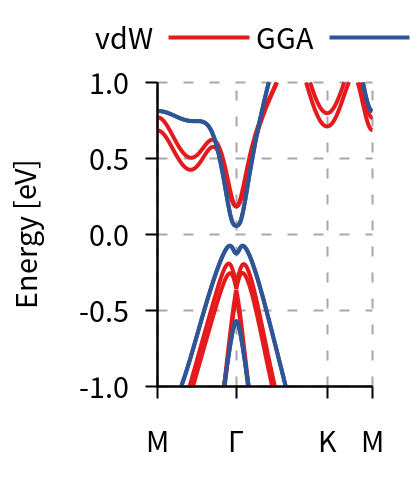}%
            \label{fig:b-hex_bands}%
        }
    \end{minipage}
    \caption{Crystal structure of freestanding, flat-hexagonal bismuthene (top view \textbf{a}; side view \textbf{b}) and buckled-hexagonal bismuthene (top view \textbf{d}; side view \textbf{e}) and corresponding DFT electronic band structure (\textbf{c} and \textbf{f}).}
    \label{fig:f-hex_b-hex}
\end{figure}

The flat-hexagonal phase of bismuthene (Fig. \ref{fig:f-hex_b-hex} a-b) refers to the arrangement of bismuth atoms in a flat honeycomb lattice (space group: P6/mmm). We relax the structure and determine the length of Bi-Bi bonds in this phase to be 3.09\,Å, which corresponds to a lattice constant of 5.35\,Å. We find that this phase of bismuthene is a semiconductor with an indirect band gap of 0.51\,eV and trivial topology ($\mathbb{Z}_2=0$).

In contrast, the buckled-hexagonal phase (Fig. \ref{fig:f-hex_b-hex} c-d) possesses a significantly smaller GGA 
band gap (0.13\,eV) and features a non-trivial topology ($\mathbb{Z}_2=1$). In this phase, the Bi-Bi bonds are slightly 
shorter (3.04\,Å) and form an angle with the lattice vectors. As a result, the atoms of the two sub-lattices are arranged in two parallel planes with a distance of 1.78\,Å (buckling height). The lattice constant shrinks to 4.28\,Å. In freestanding form, the buckled-hexagonal phase is more stable than the flat-hexagonal one ($\Delta E = 0.4\,eV$). We observe a slightly larger band gap than Singh et al. (0.08\,eV\,\cite{singh_low-energy_2019}). Considering the use of different DFT codes, pseudo-potentials and basis sets this degree of deviation in the band gap is reasonable. The band gap further increases (0.38\,eV), when we use the van der Waals (vdW) functional instead of PBE. In presence of vdW interaction the buckling height in the b-hex phase decreases (1.65\,Å) and a small in-plane magnetic moments (0.01\,$\mu_B$) emerge. 
These magnetic moments break the time-reversal symmetry and cause the splitting of the Kramer pairs (Fig.\,\ref{fig:f-hex_b-hex}f). We determined the \Z{} invariant and find that the topological phase of b-hex bismuthene is robust enough to persist despite the small spontaneous magnetization.

\subsection{SiC(0001)}

\begin{figure}[ht]
    \centering%
    \includegraphics[width=0.94\linewidth]{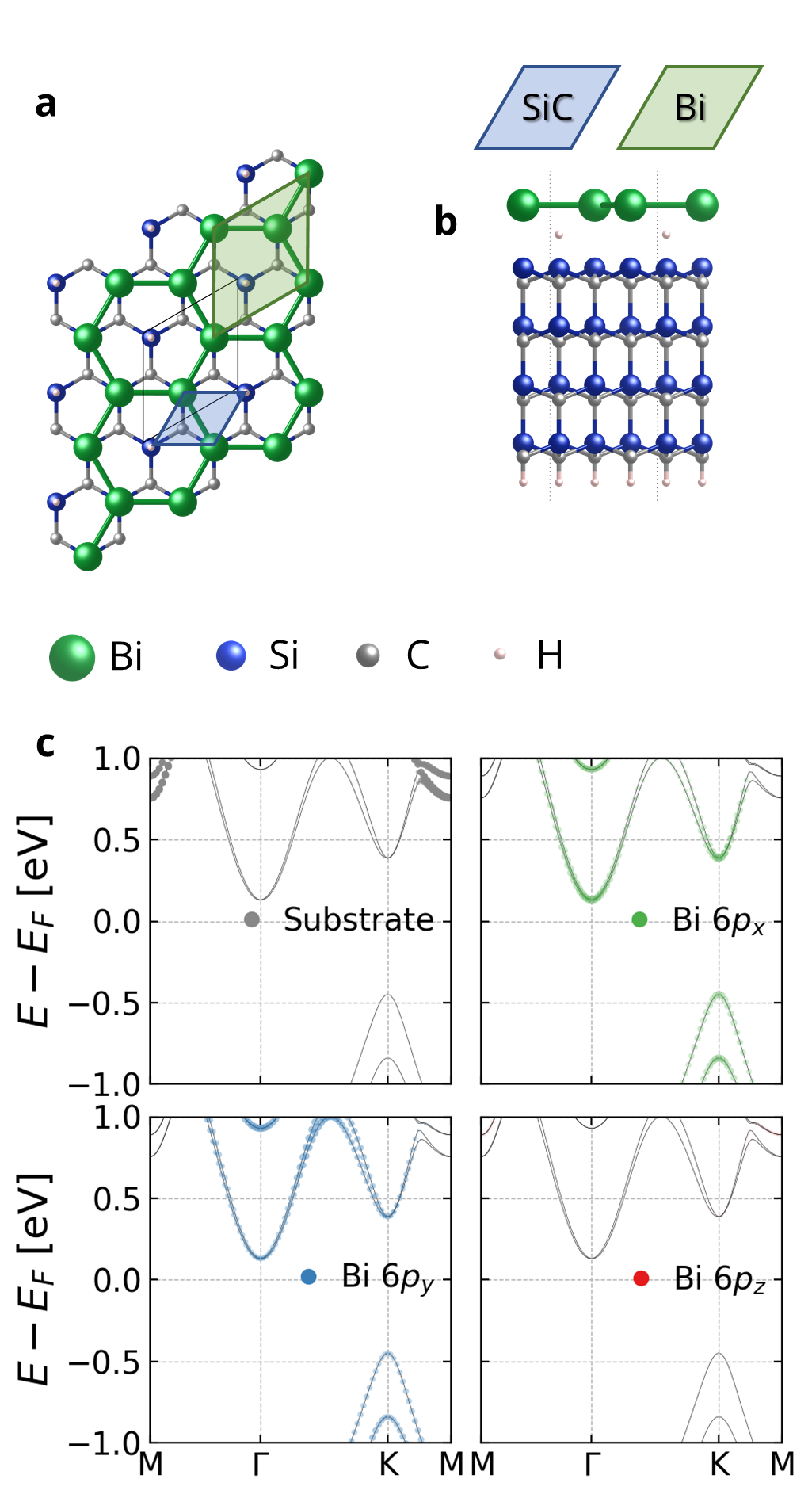}

    \caption{Crystal structure of \textbf{flat hexagonal bismuthene@SiC(0001)} (top view \textbf{a}; side view \textbf{b}). Color code and schematic of the individual unit cells are displayed in the inset. Orbital projected DFT electronic band structure (\textbf{c}): in each panel the contribution of different orbitals is proportional to the line width.}
    \label{fig:f-hex_SiC}
\end{figure}

\begin{figure}[ht]
    \centering
    \includegraphics[width=0.96\columnwidth,keepaspectratio]{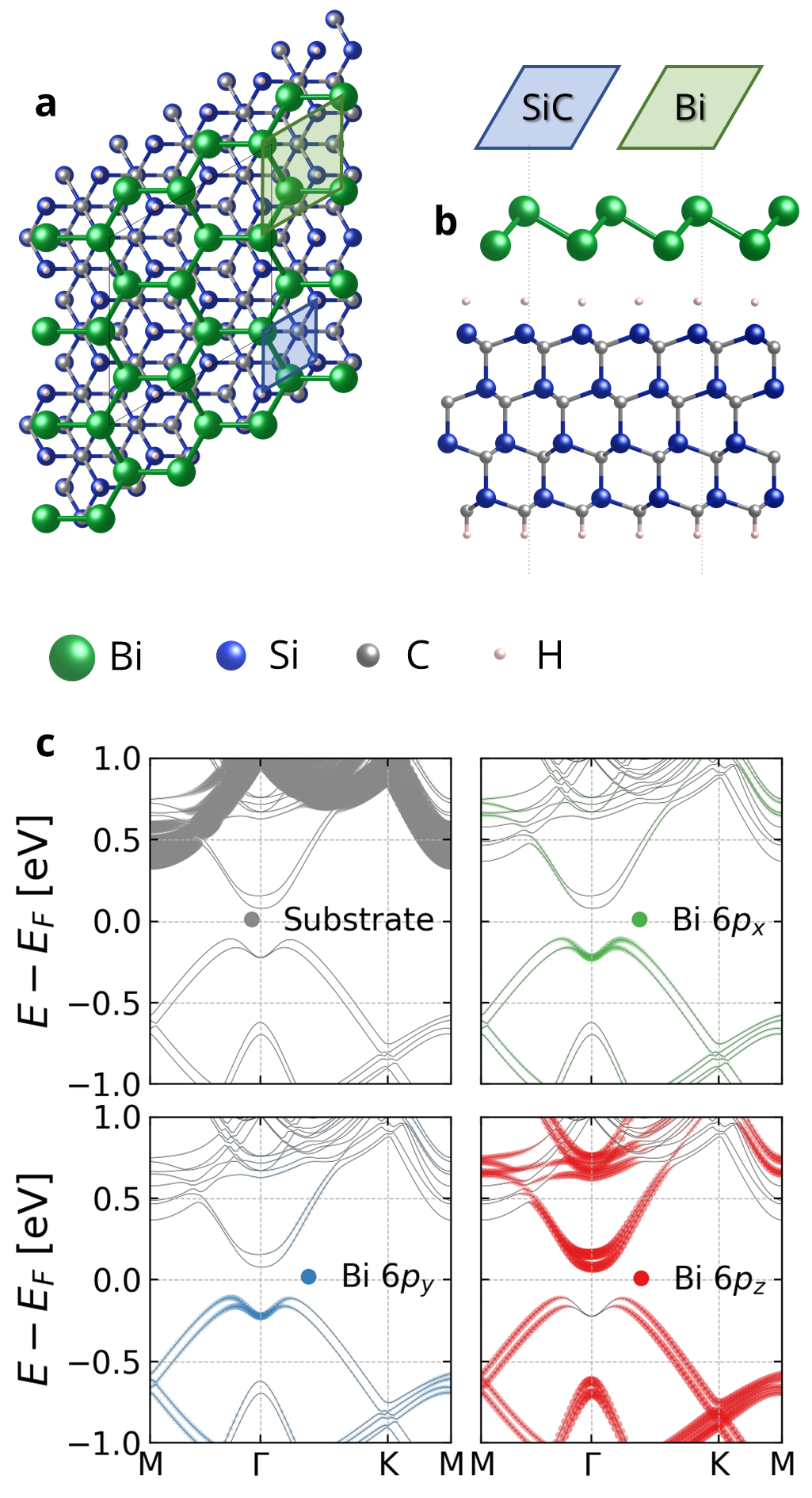}%
    \caption{Crystal structure of \textbf{buckled hexagonal bismuthene@SiC(0001)} (top view \textbf{a}; side view \textbf{b}). Color code and schematic of the individual unit cells are displayed in the inset. Orbital projected DFT electronic band structure (\textbf{c}): in each panel the contribution of different orbitals is proportional to the line width.}
    \label{fig:b-hex_SiC}
\end{figure}

In this section we show that 
both meta-stable hexagonal bismuthene phases bind to H-passivated SiC(0001) substrate, become stable (the structure could be experimentally synthesized) and are robust topological insulators.

\begin{figure}[ht]
    \includegraphics[width=0.49\textwidth]{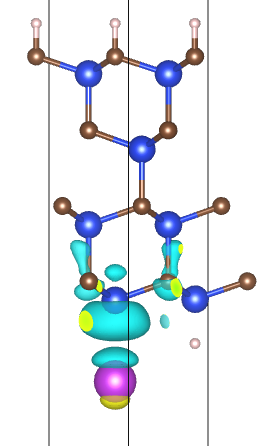}%
    \caption{\textbf{Flat hexagonal bismuthene@SiC(0001)}: Change in charge density due to interaction between monolayer and substrate. The shape of the isosurface reveals, that bonds between Bi and Si are primarily formed by $p_z$ orbitals.}
    \label{fig:f-hex_SiC_dRho}
\end{figure}

F-hex bismuthene can be grown epitaxially on SiC(0001) and becomes topologically non-trivial\,\cite{reis_bismuthene_2017}.
The \textbf{f-hex@SiC(0001)} heterostructure consists of a 1\% strained f-hex supercell on top of \RotatedUC supercell of SiC(0001) with silicon termination (Fig.\,\ref{fig:f-hex_SiC}a and b). The Bi atoms bind to the surface Si atoms leaving only the Si atom in the centre of each Bi hexagon under-coordinated
which is passivated with a hydrogen atom. The monolayer binds to the substrate with a binding energy of 1.5 eV per Bi atom.
The Bi-Si bond primarily involves $p_z$ orbitals (Fig.\,\ref{fig:f-hex_SiC_dRho})
and effectively removes the Bi-$p_z$ orbitals from the bands near the Fermi-level. 
This removal of the $p_z$ orbitals drives the phase transition from trivial to topological\,\cite{zhou_formation_2015,reis_bismuthene_2017}.

Our simulations reproduce the structural parameters and electronic structure reported by Reis et al.\,\cite{reis_bismuthene_2017}. We confirm the non-trivial topology of this combined structure and indirect band gap, with the maximum of the valence bands at K and the minimum of the conduction bands of $\Gamma$. We find only small differences in the electronic structure calculated with GGA-PBE and vdW-DRSSL functionals. For both functionals, we find a band gap of 0.6\,eV, which is slightly larger compared to Reis et al.\,\cite{reis_bismuthene_2017} (0.5\,eV). 
%

Epitaxial growth of b-hex bismuthene on SiC(0001) requires surface passivation: due to the large lattice mismatch several Si surface atoms have unsaturated bonds, which result in an unstable heterostructure. B-hex bismuthene can bind to passivated SiC(0001) via 
van der Waals interactions. The latter introduce only weak perturbations to the electronic structure and we expect it will preserve the topological phase.
A passivated surface will favor the b-hex phase or puckered monoclinic phase over the other bismuthene phases with higher free-standing formation energy 
because the van der Waals interactions are too weak to change the energetic ordering.

The \textbf{b-hex@H-SiC(0001)} heterostructure consists of a $-3$\% strained 4\ttimes 4 b-hex supercell and 3\ttimes 3 supercell of SiC(0001) with silicon termination (Fig.\,\ref{fig:b-hex_SiC}a and b).
The silicon atoms on both sides of the surface are passivated with hydrogen.
As a result the bonding between the monolayer and the substrate is weak (binding energy per Bi atom $E_B=0.08$\,eV).
The proximity interaction in b-hex@H-SiC(0001) leads to a slight deformation of the first SiC layer, despite the weak binding energy.
We computed the \Z{} invariant and found that the topological phase of b-hex bismuthene is presvered on the passivated SiC substrate.
The proximity interaction also closes the band gap (0.18\,eV) (free-standing: 0.38\,eV) and creates a more pronounced Mexican hat profile of the highest valence band.
The robustness of the topological phase and the increased band gap make SiC a reasonable candidate for the growth of b-hex bismuthene. 

\subsection{SiO$_2$(0001) $\alpha$-quartz}

    
    
    


\begin{figure*}[h]
    \centering
    \includegraphics[height=610pt]{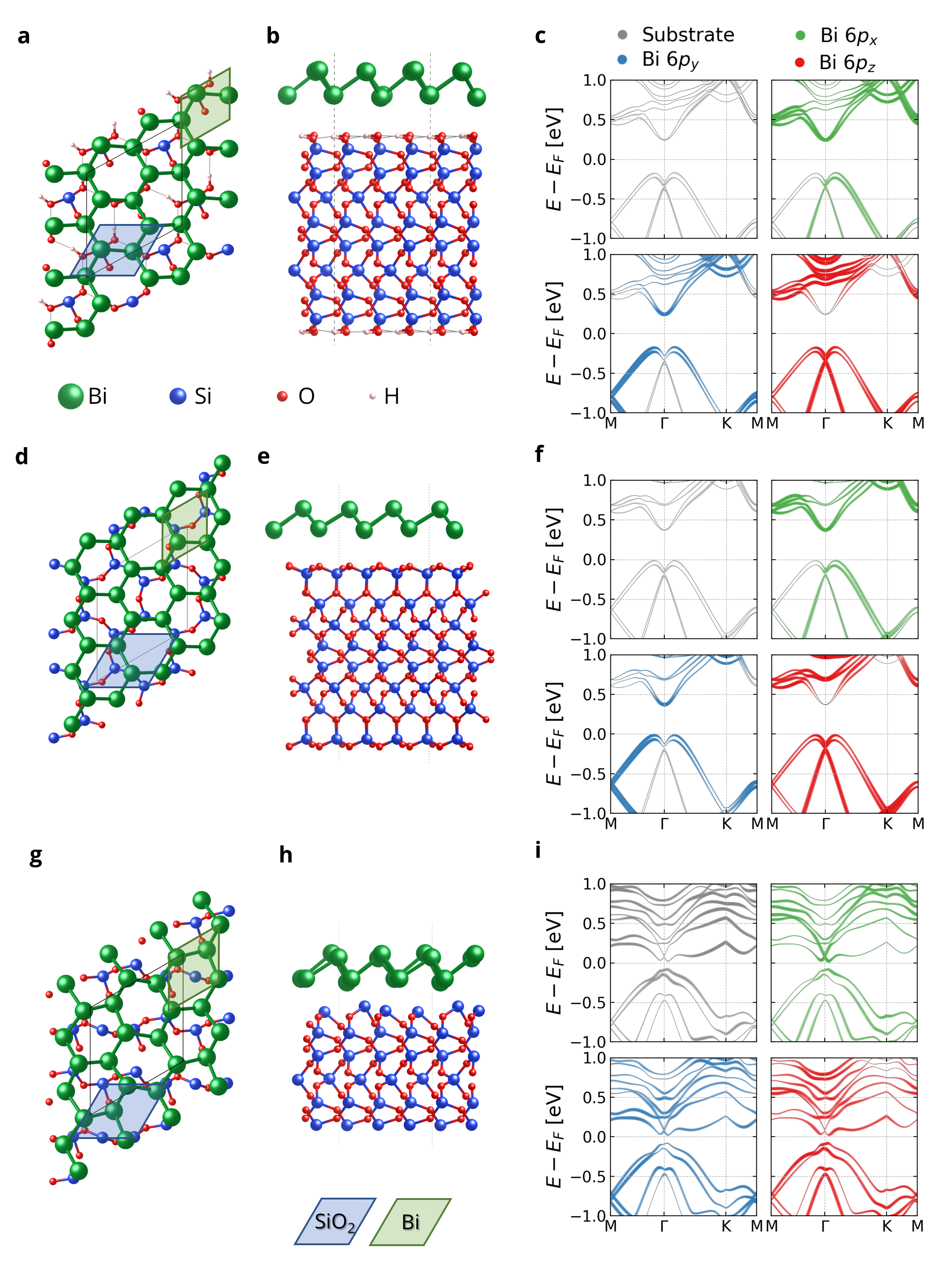}
    \caption{Crystal structure of buckled hexagonal bismuthene on SiO$_2$ (top view and side view) and orbital projected DFT electronic band structure for three different surface terminations: hydroxylated (\textbf{a-c}), reconstructed (\textbf{d-f}), and Si terminated (\textbf{g-i}). In each panel the contribution of different orbitals is proportional to the line width. On the hydroxylated and reconstructed surfaces the signature of a band inversion is visible in the projection of Bi $6p_y$ orbitals. Bi $6s$ orbitals do not contribute to the bands near the Fermi level.}
    \label{fig:b-hex_SiO2}
\end{figure*}
Silicon dioxide occurs as an amorphous material making it difficult to simulate with DFT methods which rely on periodic boundary conditions. We circumvent this limitation by simulating and comparing bismuthene monolayers on three different surfaces terminations of SiO$_2$ $\alpha$-quartz(0001): cleaved silicon terminated (SiT), reconstructed oxygen terminated (ROT), and hydroxylated silicon terminated surface (OHT) (SI Fig.~\ref{fig:b-hex_SiO2}). We omit a simply oxygen terminated surface due its structural instability. B-hex bismuthene binds to SiO$_2$, like SiC, becomes stable (could be experimentally synthesized), and the topological phase of b-hex bismuthene is preserved. However, the band gap is highly sensitive to the surface structure of the SiO$_2$ substrate. The hydroxylated, silicon terminated surface enhances the band gap, while a cleaved-silicon terminated surface surface reduced the band gap significantly. 

The b-hex@SiO$_2$ heterostructures consist of a $2\ttimes 2$ b-hex supercell on top of a \RotatedUC substrate supercell, imposing 2\% (Figure \ref{fig:b-hex_SiO2}).
B-hex bismuthene weakly binds to the three SiO$_2$ surfaces with binding energies $\leq0.11\,eV$ per bismuth atom (Table \ref{tab:summary}).

\textbf{b-hex@OHT-SiO$_2$} posses the largest band gap (0.40\,eV, Fig.\,\ref{fig:b-hex_SiO2}c)
among the three heterostructures, making it the most promising candidate for realization of quantum spin hall (QSH) states.
The band gap is increased compared to the free-standing b-hex monolayer and a rehybridization of the states near the Fermi-level occurs.
In the free-standing monolayer all three Bi-$p$ orbitals contribute equally to valence and conduction bands. In b-hex bismuthene@OHT-SiO$_2$ the symmetry D3d of freestanding b-hex bismuthene is broken (heterostructure point group is P1) lifting the $p_x$-$p_y$ symmetry: the Bi-$p_x$ and Bi-$p_y$ orbitals mainly contribute to the conduction and valence bands respectively.
As result, the valence band minimum of b-hex at $\Gamma$ is pushed to slightly higher energies and the valence bands near $M$ and $K$ are lowered (Fig.~\ref{fig:b-hex_SiO2}c) and the direct band gap at $\Gamma$ is slightly increased.
The hybridization reveals clear signature of band inversion at $\Gamma$, aside from increasing the size of the topological gap. These two characteristics indicate very robust topological phase of b-hex@OHT-SiO$_2$. We, indeed, find that the \Z{} invariant of b-hex@OHT-SiO$_2$ is 1 (non-trivial). 
The \textbf{b-hex@ROT-SiO$_2$} displays a similar behavior: the topological phase is preserved, interaction with the substrate causes a re-hybridization of the states near the Fermi level, and the signature of a band inversion is revealed (Fig.\,\ref{fig:b-hex_SiO2}f). However, band gap remains the same as in the free-standing b-hex bismuthene phase (0.38 eV).
In the \textbf{b-hex@SiT-SiO$_2$} heterostructure the spacing between monolayer and substrate is significantly smaller (Table\,\ref{tab:summary}) causing a stronger interaction between the two materials. In the relaxed structure the Bi-hexagons are skewed and buckling height varies between the atomic sites (Fig.\,\ref{fig:b-hex_SiO2}g). The proximity to the substrate also causes hybridization between Bi-$6p$ and Si-$3p$ orbitals. As a result of re-hybridization and symmetry-breaking, the band gap decreases and the splitting of the valence bands is increased
(Fig.\,\ref{fig:b-hex_SiO2}i). 
The computed \Z{} invariant shows that
the topological phase of b-hex bismuthene is retained
despite the distortion of the hexagonal structure and the hybridization with the substrate.
The reduction of the band gap of b-hex on SiT-SiO$_2$ prevents/precludes its application for quantum spin Hall devices. 
We predict that the optimal experimental realization of topological b-hex requires a 
hydroxalated SiO$_2$ surface. 

F-hex bismuthene also binds to the SiO$_2$ surfaces: weakly to the OHT-SiO$_2$ and ROT-SiO$_2$ surfaces, strongly to SiT-SiO$_2$, The latter distorts the monolayer and breaks its symmetry. The topologically trivial phase of f-hex monolayer is retained on all three surfaces. The computed band structures are reported in the SI (SI Fig.\,8).

\subsection{Si(111)}
The topologically non-trivial phase of b-hex bismuthene binds to Si(111) and becomes stable. However, the proximity interaction slighly lowers the band gap, making Si(111) a poor choice for topological insulator devices based on b-hex bismuthene. 

\begin{figure}
    \includegraphics[width=0.96\columnwidth]{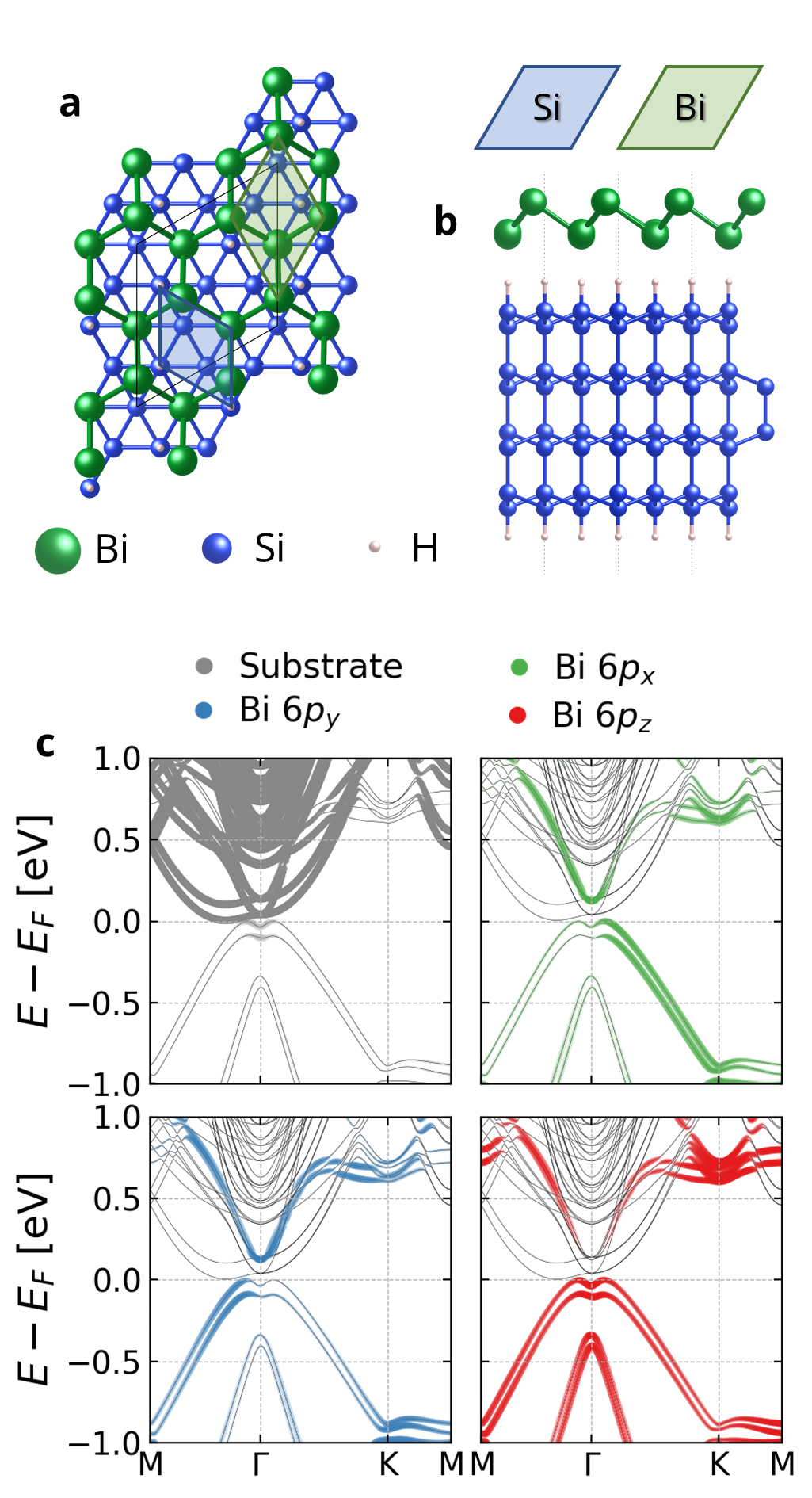}%
    \caption{Crystal structure of \textbf{buckled hexagonal bismuthene@passivated Si(111)} (top view \textbf{a}; side view \textbf{b}). DFT electronic band structure (\textbf{c}) in each panel the contribution of different orbitals is proportional to the line width.}
    \label{fig:b-hex_Si+H}
\end{figure}

The b-hex on SI(111) heterostructure consists of a 4\% strained \RotatedUC b-hex supercell and 2\ttimes 2 Si(111) supercell (Fig.\,\ref{fig:b-hex_Si+H}a and b), with a binding energy of 0.07\,eV per Bi atom. 
Passivation of the Si(111) substrate is required to prevent the heterostructure from being metallic. 
Similar to the b-hex@SiO$_2$ heterostructures the Si(111) substrate breaks the symmetry of the monolayer and as result the orbital contribution of Bi-p$_x$ and Bi-p$_y$ to the bands are not symmetric. In this heterostructure 
no hybridization between the monolayer and the substrate occurs and the electronic structures of the two components are largely independent. 
The Fermi-levels in this heterostructure are aligned such that the valence band minimum of bismuthene lays just a bove the conduction band minimum, and the band gap closes.
However, the topological band gap should not be measured between the conduction band minimum of the substrate and the top of the Bi valence, because the independence of the components. Rather, the relevant gap for realization of quantum spin hall states should be measured between Bi bands: 0.12\,eV (Fig.\,\ref{fig:b-hex_Si+H}c). This gap is reduced significantly with respect to the free-standing phase making Si a poor choice for b-hex bisumuthene substrate. 

\section{Conclusion}
In this work, we studied the proximity interaction between hexagonal Bi monolayer phases and silicon-based substrates to identify potential substrates for room-temperature TI applications based on Bi monolayers.
We demonstrate that buckled-hexagonal and flat-hexagonal bismuthene can be stabilized on H-SiC(0001), H-Si(111) and on $\alpha$-quartz SiO$_2$(0001).
We show that the proximity interaction in the heterostructure has significant effect on the electronic structure of the monolayer, even when no bonding occurs.
We further identify the structural realignment and the breaking of the sublattice symmetry of b-hex monolayer as the main factors driving the change in the electronic structure. The exact nature of this proximity interaction is sensitive to the substrate material and its surface composition.
In particular, the size of the topological gap of b-hex bismuthene varies drastically based on the chosen substrate.
Importantly, the topology of the hexagonal Bi monolayers (f-hex: trivial, b-hex: non-trivial) is unaffected by the interaction with the substrate, with the exception of f-hex@SiC(0001).
Hydroxylated SiO$_2$ emerges as an especially good substrate choice for b-hex bismuthene due to an increase of the topological gap and could enable room-temperature application.

\section{\textit{Acknowledgments}}

ZZ acknowledges financial support by the Ramon y Cajal program (RYC-2016-19344), and the Netherlands Sector Plan program 2019–2023.
ICN2 is supported by the Severo Ochoa program from Spanish MINECO (Grant No. SEV-2017-0706) and the CERCA Program of Generalitat de Catalunya. PO acknowledges support from Spanish MICIU, AEI and EU FEDER (Grant No. PGC2018-096955-B-C43) and the European Union MaX Center of Excellence (EU-H2020 Grant No. 824143).
We acknowledge computing resources on MareNostrum4 at Barcelona Supercomputing Center (BSC), provided through the PRACE Project Access (OptoSpin project 2020225411) and RES (activity FI-2020-1-0014), resources of SURFsara the on National Supercomputer Snellius (EINF-1858 project) and technical support provided by the Barcelona Supercomputing Center.

NW acknowledges support from the European Union’s Horizon 2020 research and innovation programme under the Marie Skłodowska-Curie grant agreement No. 754558.


\bibliography{Bibliography}

\end{document}